\documentstyle[11pt,newpasp,twoside,psfig]{article}

\def\edcomment#1{\iffalse\marginpar{\raggedright\sl#1\/}\else\relax\fi}
\reversemarginpar

\newcommand{\chandra}{{\em Chandra}}
\newcommand{\XMM}{{\em XMM}}
\newcommand{\etal}{\mbox{{\em et~al.}}}
\newcommand{\NH}{\mbox {$N_{\rm H}$}}

\begin{document}

\title{The Low-Energy Spectral Response of the ACIS CCDs}
\author{Paul P. Plucinsky, Richard J. Edgar and Shanil N. Virani}
\affil{Harvard-Smithsonian Center for Astrophysics, 60 Garden St.,
MS-70, Cambridge, MA 02138}

\author{Leisa K. Townsley and Patrick S. Broos}
\affil{Department of Astronomy \& Astrophysics, The Pennsylvania State
University, 525 Davey Lab, University Park, PA 16802}


\begin{abstract}
The flight calibration of the spectral response of the 
{\em Advanced CCD Imaging Spectrometer} (ACIS) below 1.5
keV is difficult because of the lack of strong lines in the
calibration source on-board the {\em Chandra X-ray Observatory}.
We have been using 1E0102.2-7219 (the brightest supernova remnant in the SMC)
to evaluate the ACIS response matrices, since this remnant has strong
lines of O, Ne and Mg below 1.5 keV. The spectrum of 1E0102.2-7219 has
been well-characterized using the gratings on \chandra\/ and \XMM\/.
We have used the high-resolution spectral data from both gratings
instruments to develop a spectral model for the CCD spectra.  
Fits with this model are sensitive to any problems with  
the gain calibration and the spectral redistribution model.
We have fit the data with the latest response matrix 
for the S3 device released
in August 2001 (available in CALDB~2.7 or higher).
We have also applied the charge-transfer inefficiency (CTI) correction
software (version 1.37) developed at Penn State to both the
Backside-illuminated (BI) and Front-side illuminated (FI) data
and fit the data with the associated CTI-corrected response matrices.
All fits show a significant
improvement in the low energy response with the new matrices.
The fits to the FI CCDs demonstrate the utility of the FI CCDs for spectral
analysis, in spite of the radiation damage the devices suffered early
in the mission.
\end{abstract}

\vspace{-0.25in}

\section{Introduction}

   The {\em Advanced CCD Imaging Spectrometer} (ACIS) instrument team, 
at both Penn State and MIT, and the
Chandra X-ray Center (CXC) have been working to improve the quality of the
flight calibration of the ACIS CCDs.  The CXC has recently released a
new response matrix for the S3 CCD at -120~C in the CALDB 2.7 and
CIAO 2.1.3 release (see
asc.harvard.edu/cal/Links/Acis/acis/Cal\_prods/ \\
matrix/matrix.html).  
The new matrix includes a more sophisticated representation of the
spectral redistribution function and an improved model of the gain
at energies below 1~keV.  The major difference between the CALDB 2.7
matrices and previous matrices is at energies below 1~keV, with little
or no difference at energies above 1.5~keV. There has been no change
to the quantum efficiency model of the detector. 

The ACIS instrument team has released software (SW) to correct for 
the charge-transfer inefficiency (CTI)
of both the Frontside-illuminated (FI) and Backside-illuminated (BI)
CCDs.  The physical model of the CCD is described in Townsley~\etal\/
(2001a) and the details of the CTI correction are given in
Townsley~\etal\/~(2000) and
Townsley~\etal\/~(2001b).  The CTI correction SW (version 1.37) 
and accompanying response matrices (30JUL01)  are available at
``www.astro.psu.edu/users/townsley/cti/''
or from the contributed SW page at the CXC web site
``asc.harvard.edu/cgi-gen/cont-soft/soft-list.cgi''.  The CTI
correction for the BI CCD is small but is rather large for the FI
CCDs, and makes a dramatic difference in spectral fits with the FI CCDs.

  In this paper, we will focus on the low energy response as described
by the response matrices listed above. Specifically, we will use
strong lines in the spectrum of the bright, SMC supernova remnant
(SNR) 1E0102.2-7219 in the
region between 0.5-1.5~keV to test the new response matrices.  The
spectrum of 1E0102.2-7219 has been well-characterized using the gratings on
\XMM\/ (see Rasmussen~\etal\/ 2001 and Sasaki~\etal\/~2001) and also
on \chandra\/ (see Canizares~\etal\/~2001, Flanagan~\etal\/~2001, and
Davis~\etal\/~2001).  The spectrum is dominated by strong lines from
O~VII, O~VIII, Ne~IX, Ne~X, Mg~XI, and Mg~XII.  The high resolution
gratings spectra show that there is little or no Fe emission in this
spectrum. The lack of Fe in the 1E0102.2-7219 spectrum greatly
simplifies the modeling of the spectrum, since the Ne lines are not
blended with the Fe~L lines.




\vspace{-0.25in}
\section{Data}

  We used six datasets for the S3(BI) analysis and five datasets for the
I3(FI) analysis.  The datasets are listed in Table~1 with the date of
observation, position on the CCD, and exposure time.  The six datasets
for S3 sample four regions on the CCD, since two of the observations
were executed at the same location on the chip but at different times
to check for temporally-varying performance.  The five datasets on I3
sampled five different locations in row number on the CCD in order to
characterize the effect of CTI on low-energy X-rays.  All of these
observations are calibration observations and are available in the 
{\chandra} public archive.

\begin{table}
\centering
\caption{{\bf List of 1E0102.2-7219 Observations Used}}

\begin{tabular}{|r|l|c|c|l|l|l|}

\hline
OBSID & DATE        & y-offset & z-offset & CCD & CCD &
Exp  \\
      &            &  (arcmin) & (arcmin) & column & row  &  
 (ks) \\
\hline
\multicolumn{2}{l}{\bf BI(S3) Datasets } & & & & &    \\
\hline
 1311 & 2000-12-10  & -1.0      & 0.0       & 374   & 506   & 7.8  \\
 1531 & 2001-06-06  & -1.0      & 0.0       & 362   & 507   & 7.5 \\
 1308 & 2000-12-10  &  1.0      & 0.0       & 130   & 508   & 8.0 \\
 1530 & 2001-06-06  &  1.0      & 0.0       & 116   & 508   & 7.7 \\
  141 & 2000-05-28  & -1.0      & 2.0       & 364   & 262   & 9.8 \\
 1702 & 2000-05-28  & -1.0      & -2.0      & 366   & 752   & 9.6 \\
\hline
\multicolumn{2}{c}{\bf FI(I3) Datasets } & & & & &  \\
\hline
  440 & 2000-04-04  & -0.5      & 2.5       & 662   & 904   & 6.9  \\
  439 & 2000-04-04  & -2.25     & 2.5       & 662   & 690   & 6.9 \\
  136 & 2000-03-16  & -4.0      & 2.5       & 661   & 476   & 10.0\\
  140 & 2000-04-04  & -5.5      & 2.5       & 662   & 291   & 8.3 \\
  420 & 2000-03-14  & -7.0      & 2.5       & 658   & 104   & 10.2
\\
\hline

\end{tabular}

\end{table}

 The spectra were extracted and response matrices and auxiliary response
files generated using the standard {\it CIAO} tools {\tt dmextract, mkrmf,
mkarf} using {\it CIAO} 2.1.3 and CALDB 2.7 for the S3 data.  For the I3
data and also for the S3 dataset OBSID 1311, the spectra were
extracted using {\tt dmextract} but specifying
the binning of PI channels described in the recipe on the CTI
correction SW page. The different channel types are labeled in
Figure~1 as `` PHA'' and ``PI'' for the channel types produced by
the {\it CIAO} SW and ``PPI'' for the channel types produced by the
Penn State CTI-correction SW.
 We used the response matrices for I3 and S3 appropriate 
for CTI-corrected data based on the position of the source on the chip
(see the CTI-correction web pages for more details).  We used an
annulus for the spatial extraction region in order to maximize the
contribution of the shocked ejecta and to minimize the contribution of
the primary blastwave of the remnant (see Hughes~\etal\/ 2001 and
Gaetz~\etal\/ 2001).  Extracting spectra from this annulus should 
maximize the contrast  between the lines and the underlying continuum.
%


\section{Spectral Analysis}

\subsection{Spectral Model}

  We developed a spectral model (based on the high-resolution gratings
data) which consisted of two components for absorption (one for the
galactic contribution and one for the SMC contribution with variable
abundances), a bremsstrah\-lung component for the continuum, and
Gaussians  for the
lines.  The energies of the Gaussians were set to their known values
and not allowed to vary during the fitting process.  The widths of the
lines were set to zero, so that the width is completely determined by the
spectral redistribution function of the matrix. 
The HETG data indicate
that there are measurable Doppler shifts in these lines on the order
of a few~eV, but these are small compared to the resolution of the detector
($\sim75-150$~eV).
Only the
normalizations of the lines were allowed to vary.  In this manner, this
model was used to verify that the lines were modeled at the correct
energy in the matrices and the shapes of the lines were consistent
with the  widths contained in the matrices.  
Our model contains a total of 24 lines determined from the 
gratings spectra, but in most of the fits three or four of  the lines are
fitted with a zero normalization.

\subsection{S3(BI) Spectral Fits}

The spectral model described above was fit to all of the datasets in
Table~1.  The values of the reduced $\chi^2$ ranged from 1.1 to 2.1.
Table~2 and Figure~1 summarize the fitted values for the \NH\/ and $kT$.
OBSID 1311 was fitted in both PHA and PI channel space with the
CALDB~2.7 matrices and the resulting values agreed well with each
other.  OBSID~1311 was also fitted after applying the CTI correction.
The fitted value of $kT$ was consistent with the value derived from the
CXC matrix fits, but the fitted value of the \NH\/ was discrepant at 
the 90\%\/ confidence limit (CL).  OBSIDs 1531, 1308, 1530, 141, and
1702 were fitted in PHA space and compared to OBSID~1311.  All
datasets agreed within the 90\%\/  CL except for
OBSID~1530. OBSID~1308 agreed within the 90\%\/ CL, but had the 
largest error bars by far of any dataset.  Both OBSID~1530 and 1308
warrant further investigation for instrumental effects which may have
compromised the data.


\begin{figure}

\hbox{

\vbox{\psfig{figure=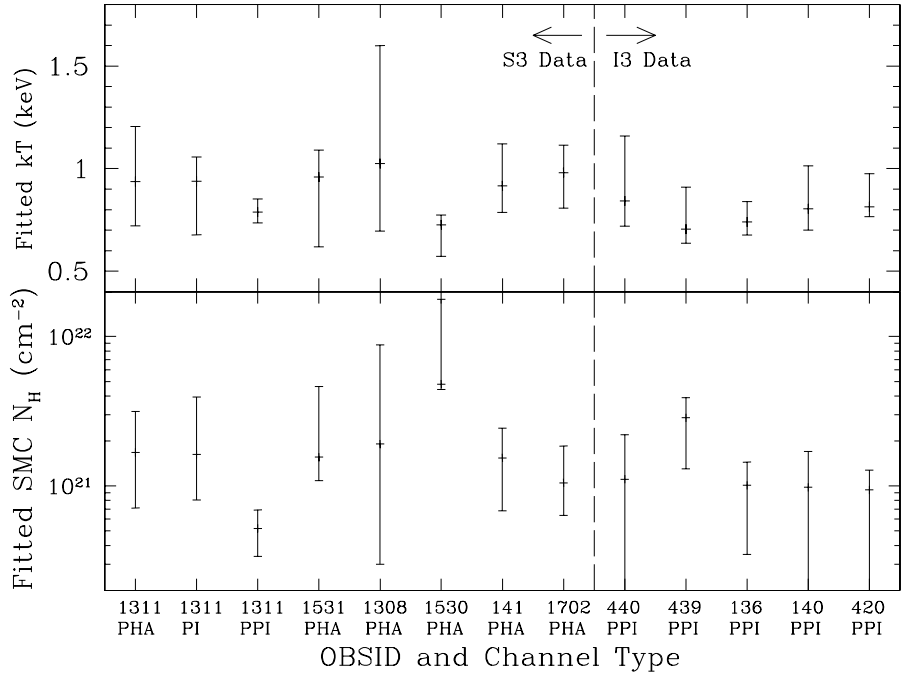,width=5.0in}}

}

\caption{Fitted Values vs OBSIDs for S3(BI) and I3(FI).  The channel
types ``PHA'' and ``PI'' refer to channels produced by the {\it CIAO}
SW and ``PPI'' refers to channels produced by the Penn State CTI
correction SW. Error bars are 90\% CL.}

\end{figure}

Figure~2 displays the fit to OBSID~1311 in PHA space.  The matrix and
model represent the data extremely well from 700~eV to 1600~eV.  Note
how well the Ne lines at $\sim900$~eV and $\sim1000$~eV are fitted by
this matrix and model.  There are data from 1.6~keV to 5.0~keV which we 
have chosen to exclude from the plot since there are no strong lines
in that region, but which are included in the fit.  Only below 700~eV
is there any evidence of systematic effects in the residuals.  The
fits to the other data sets and to the CTI-corrected data are
qualitatively similar.  These data confirm that the CALDB~2.7
matrix has correctly incorporated the gain and spectral redistribution
function of the BI(S3) CCD and that data from four different locations on
the S3 detector produce similar fitted results.


\begin{table}[tbp]
\centering

\caption[ ]{{\bf {Spectral Fit Results with 90\%\/ Confidence Limits }}}

\begin{tabular}{|r|l|l|l|l|}
\hline
OBSID & SMC \NH\/($10^{21}~{\rm cm^{-2}}$) & $kT$(keV) & Red
$\chi^2$ & DOF  \\
\hline
\multicolumn{2}{l}{\bf BI(S3) Datasets } & & &    \\
\hline
 1311 & 1.68 [0.71,3.15]   & 0.94 [0.72,1.21] & 1.46 & 117     \\
 1311 & 1.63 [0.80,3.94]   & 0.94 [0.68,1.06] & 1.73 & 78 (CXC PI)     \\
 1311 & 0.52 [0.34,0.69]   & 0.79 [0.74,0.85] & 1.35 & 80 (PSU PI)    \\
 1531 & 1.56 [1.08,4.63]   & 0.96 [0.62,1.09] & 1.31 & 111     \\
 1308 & 1.91 [0.30,8.80]   & 1.03 [0.70,1.60] & 2.06 & 121 \\
 1530 & 4.80 [4.43,17.8]   & 0.73 [0.57,0.77] & 1.46 & 118 \\
  141 & 1.54 [0.68,2.43]   & 0.92 [0.79,1.12] & 1.68 & 125  \\
 1702 & 1.05 [0.64,1.85]   & 0.98 [0.81,1.11] & 1.17 & 125 \\
\hline
\multicolumn{4}{l}{\bf FI(I3) Datasets, CTI-Corrected, PSU matrix }  &  \\
\hline
  440 & 1.11 [0.10,2.20]   & 0.84 [0.72,1.16] & 1.28  & 59  \\
  439 & 2.86 [1.30,3.90]   & 0.71 [0.64,0.91] & 1.08  & 61  \\
  136 & 1.01 [0.35,1.44]   & 0.74 [0.68,0.84] & 1.52  & 63  \\
  140 & 0.98 [0.00,1.71]   & 0.80 [0.70,1.10] & 1.74  & 58  \\
  420 & 0.94 [0.00,1.26]   & 0.81 [0.77,0.98] & 1.89  & 60  \\
\hline

\end{tabular}

\end{table}

\begin{figure}

\hbox{

\vbox{\psfig{figure=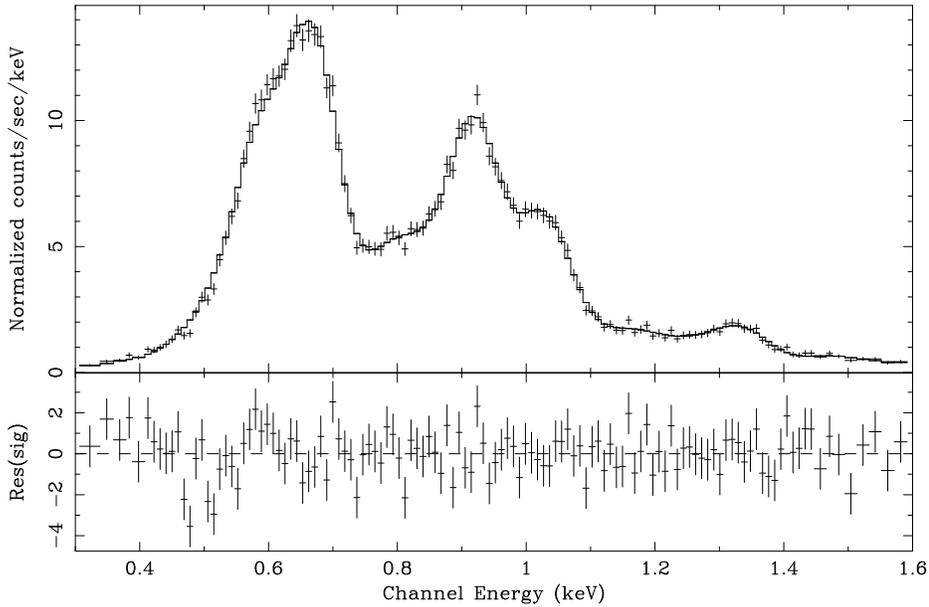,width=5.0in}}
}

\caption{OBSID 1311: Fit with the CXC CALDB 2.7 S3 matrix in PHA space. }

\end{figure}

\subsection{I3(FI) Spectral Fits}

We fit the five datasets on the I3 CCD which span the regions from low
row numbers where the CTI effects are minimized to high row
numbers where the CTI effects are largest. The results are summarized
for OBSIDs~440, 439, 136, 140, and 420 in Table~2 and Figure~1.  The
fitted values for the \NH\/ and $kT$ are all consistent with each other
at the 90\%~CL, except for the \NH\/ for OBSIDs 420 and 439.  The
consistency of the fitted results demonstrates the value and accuracy
of the CTI correction.  The spectra for OBSIDs~420, 136,
and 440 are shown in Figures~3, 4,and 5.  The quality of the fits
is good with the reduced~$\chi^2$ values ranging from 1.1 to 1.9.
Only around 650~eV is there any 
evidence of systematic effects in the residuals.
It is clear from the these
plots that the spectral resolution of the detector changes
dramatically with row number.  The gain of the detectors is
also changing, but this effect has also been corrected by the CTI
correction SW.  The data from OBSID 420 show that the spectral
resolution of I3 at low row numbers is significantly better than S3
(see Virani~\etal\/ these proceedings).  Note how cleanly the O and Ne
lines are separated in Figure~3 and compare to the S3 spectrum in
Figure~2.  In fact, OBSID~136 shows that I3 has better spectral
resolution than S3 halfway across the CCD.  It is only in the top half
of the FI CCDs where the resolution is worse than S3 after correcting 
for CTI as shown in Figure~5.




\begin{figure}

\hbox{

\vbox{\psfig{figure=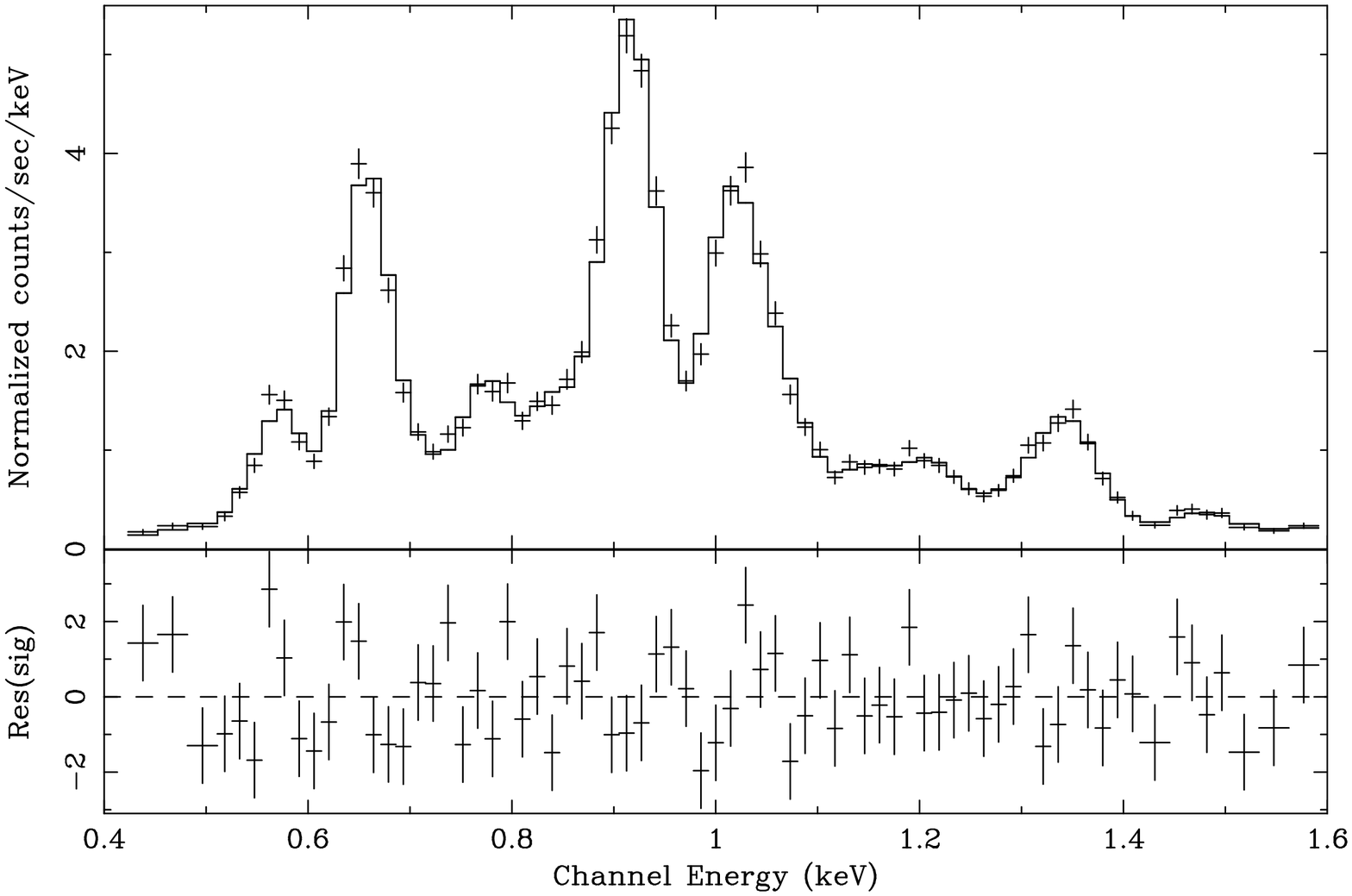,width=5.0in}}
}

\caption{OBSID 420: Fit with the PSU 30JUL01 matrix in PI space (row=104). }

\end{figure}


\begin{figure}

\hbox{

\vbox{\psfig{figure=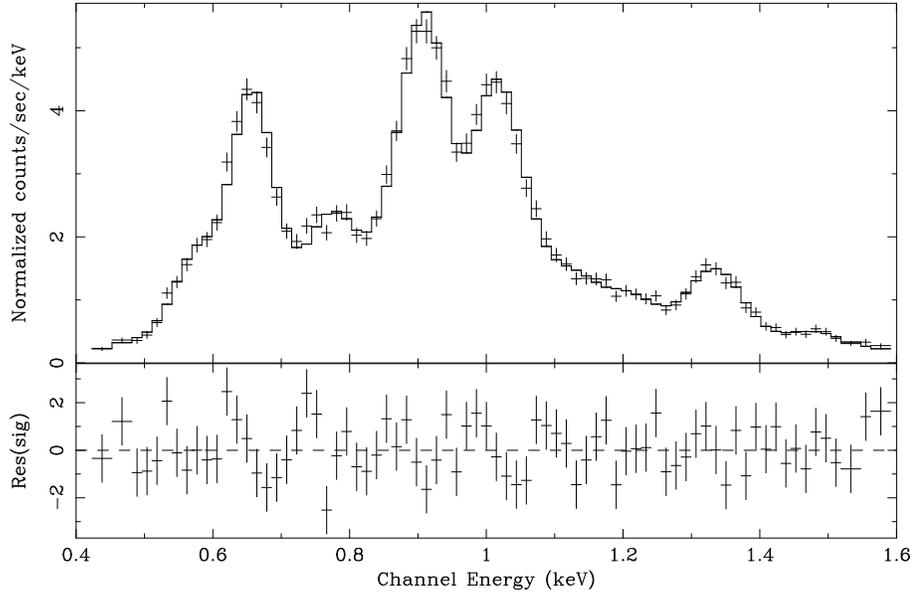,width=4.9in}}
}

\caption{OBSID 136: Fit with the PSU 30JUL01 matrix in PI space (row=476). }

\end{figure}


\begin{figure}

\hbox{

\vbox{\psfig{figure=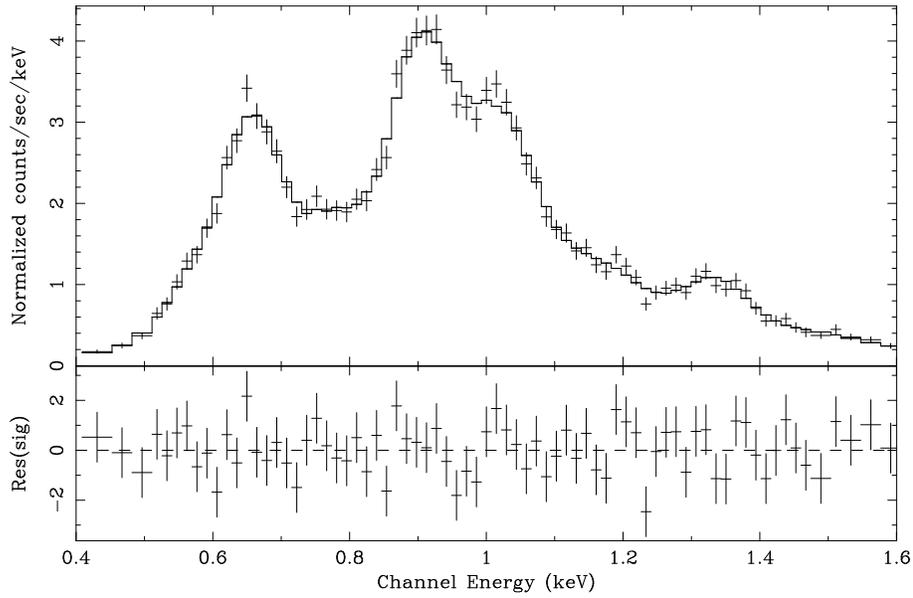,width=4.9in}}
}

\caption{OBSID 440: Fit with the PSU 30JUL01 matrix in PI space (row=904). }

\end{figure}

\vspace{-0.15in}
\section{Conclusions}

We have used the soft, line-dominated spectrum of the SMC SNR 
1E0102.2-7219 to evaluate the response matrices of the BI and FI
CCDs at low energies.  We have demonstrated that the new BI(S3)
matrices released in August 2001 in CALDB~2.7 and in the Penn State
CTI correction SW are an accurate representation of the spectral
response of the S3 CCD from 0.7 to 1.6~keV.  We have used several
datasets of 1E0102.2-7219 on S3 to verify that the new matrix yields
consistent fitted parameters for different locations on the CCD.
We have applied the Penn State CTI correction SW to five datasets
on the I3(FI) CCD and demonstrated that the SW corrects for the
spatial dependence of the gain and the spatial dependence of the
redistribution function due to CTI.  The fits indicate that the
response matrix is an accurate representation of the spectral response
of the I3 CCD.  We derive fitted parameters which are consistent with
each other at the 90\%\/ CL, even though  the spectral resolution of
the I3 CCD is changing rapidly with position.  In fact, after CTI
correction the FI CCDs have superior spectral resolution compared to
S3 across the bottom half of the CCDs.  \chandra\/ observers should be
aware of this performance and use it to their advantage if possible.

\vspace{-0.15in}
\section{Acknowledgments}

We thank all of our colleagues at the CXC, PSU, and MIT who have
helped with the flight calibration of the ACIS instrument.  We thank
all of the engineers, technicians, and scientists who have made the
{\em Chandra X-ray Observatory} such a success. PPP, RJE, and SNV 
acknowledge support for this work from NASA contract NAS8-39703.
LKT and PSB acknowledge support for this work from NASA contract
NAS8-38252.

\end{document}